\documentclass[aps,nofootinbib,showpacs,preprintnumbers,amsmath,amssymb]{revtex4}

\def\be{\begin{equation}}
\def\ee{\end{equation}}
\def\ba{\begin{eqnarray}}
\def\ea{\end{eqnarray}}
\newcommand{\A}{\mathcal{A}}

\newcommand{\tlA}{\mathfrak A}
\usepackage{graphics}
\usepackage{graphicx}% Include figure files
\usepackage{dcolumn}% Align table columns on decimal point
\usepackage{bm}% bold math
\usepackage{epsfig}
\usepackage{graphicx}
\begin{document}

\preprint{USM-TH-253; appeared in JPG37,075001}

\title{Analytic QCD coupling with no power terms in UV regime}

\author{Gorazd Cveti\v{c}}
 \email{gorazd.cvetic@usm.cl}
\affiliation{Department of Physics, Universidad T{\'e}cnica Federico
Santa Mar{\'\i}a, Valpara{\'\i}so, Chile}
\affiliation{Center of Subatomic Studies and Scientific-Technological Center of Valparaiso, Chile}

\author{Reinhart K\"ogerler}
\email{koeg@physik.uni-bielefeld.de}
\affiliation{Department of Physics, Universit\"at Bielefeld, 33501 Bielefeld, Germany}

\author{Cristi\'an Valenzuela}
 \email{cvalenzuela@fis.puc.cl}
\affiliation{Department of Physics, Pontificia Universidad Cat\'olica de Chile, Santiago 22, Chile}

\date{\today}

\begin{abstract}
We construct models of analytic QCD (i.e.,with the running coupling parameter
free of Landau singularities) which address several problems encountered
in previous analytic QCD models, among them their incompatibility
with the ITEP-OPE philosophy (due to UV power terms) and too low values of the
semihadronic $\tau$ decay ratio. The starting point of the approach is
the construction of appropriate nonperturbative beta functions.
\end{abstract}

\maketitle

\section{Introduction}
\label{sec:intr}

In the perturbative QCD (pQCD), the running coupling
$a_{\rm pt}(Q^2) \equiv \alpha_{s}(Q^2)_{\rm pt}/\pi$,
as a function of $Q^2 \equiv -q^2$ ($q$ being a typical
four-momentum transfer of the considered physical process),
has so called Landau singularities at $0 < Q^2 \leq \Lambda^2$
($\Lambda^2 \sim 10^{-1} \ {\rm GeV}^2$),
thus not reflecting the analyticity of the
space-like observables ${\cal D}(Q^2)$ for all
$Q^2 \in  \mathbb{C} \backslash (-\infty,0]$ dictated by the
causality of quantum field theories. As a consequence,
the evaluated expressions of such observables in pQCD have
wrong analyticity properties and become unreliable
at low $|Q^2|$. In order to overcome this fundamental problem of
pQCD, several attempts have been made to restore the correct
analytic (i.e., holomorphic) properties of both the
coupling parameter $a(Q^2)$ and the related evaluated 
expressions of observables, which all go under the generic name
of analytic QCD (anQCD).

Various models of anQCD found in the literature
(some of them: Refs.~\cite{ShS}-\cite{CV2};
for reviews and further references see 
\cite{Prosperi:2006hx}-\cite{Cvetic:2008bn}), among them
the most popular minimal analytic (MA) model of Shirkov and Solovtsov 
\cite{ShS}, have faced criticism based mainly on one or both of the
following points, one being theoretical and the other phenomenological:
\begin{enumerate}
\item
The analytic running coupling parameter $a(Q^2)$
differs at large $Q^2$ from the ordinary pQCD coupling
$a_{\rm pt}(Q^2) \equiv \alpha_s(Q^2)/\pi$ 
by power terms $\sim (\Lambda^2/Q^2)^n$ or  
$\sim (\Lambda^2/Q^2)^n \ln^m (\Lambda^2/Q^2)$
[in MA: by terms $\sim (\Lambda^2/Q^2)^1$].
However, these then lead in (inclusive) physical observables
to the corresponding power corrections which, nota bene,
come from the ultraviolet (UV) regime \cite{Cvetic:2007ad}.
If such observables are calculated
within the operator product expansion (OPE) mechanism, it
is readily seen that such power terms are in conceptual 
contradiction with the general OPE philosophy which has been
vigorously advocated in particular by the ITEP group 
 \cite{Shifman:1978bx} (see also, e.g., Ref~\cite{DMW}) and whose validity
is strongly indicated by the success of the related QCD sum rules.
This philosophy rests on the assumption that OPE is true in
general (not only in the perturbative approach) and that it allows
to separate short-range contributions from the long-range ones.
And it is only the long-range contributions which should lead to power
corrections, reflecting the nonperturbative physics. Thus, there is
no space for UV-generated power corrections within the ITEP-OPE
philosophy.
\item In the widely
used MA model, the prediction for one of the best-measured low-energy
QCD observables, namely the strangeless $r_{\tau}$, the decay branching
ratio of the $\tau$ lepton into nonstrange hadrons, lies in the
region $r_{\tau} \approx 0.14$ \cite{Milton:1997mi}, significantly below
the experimental value $r_{\tau} = 0.202 \pm 0.004$. It appears that
anQCD models in general tend to give too low values of  
$r_{\tau}$ \cite{Geshkenbein:2001mn}.
%orders of magnitude than their current mass values \cite{Milton:2001mq});
\end{enumerate}

Point (2) (the $r_{\tau}$ problem) can be addressed by introducing 
additional parameters (Refs.~\cite{Milton:2001mq,CV2}); in MA,
these parameters (the quark masses) have to be chosen unusually large
\cite{Milton:2001mq}. On the other hand, point (1) (ITEP-OPE) has 
not been addressed in a more systematic way
in the literature hitherto.\footnote{In Ref.~\cite{Alekseev:2005he}
the problem is addressed in an approximate way, by requiring the 
aforementioned power index $n$ to be large ($n=4$). In 
Ref.~\cite{Cvetic:2007ad} it was explored whether an analytic coupling 
respecting ITEP-OPE can be constructed directly; it turned out
to be difficult, and several parameters had to be introduced.}
Within this letter we try to develop a version of analytic QCD which
addresses both problems mentioned above. We base our approach on
the assumption that the singularity structure of $a(Q^2)$
reflects the singularity structure of space-like observables
${\cal D}(Q^2)$ in a ``minimal'' way, i.e., dictated by physical
principles (causality and unitarity) which means that $a(Q^2)$ is
analytic in the complex $Q^2$-plane with the exception of a cut along
the negative semiaxis starting at $Q^2 = - M_{\rm thr}^2 < 0$
(there are no massless hadrons). Specifically, we expect
$a(Q^2)$ to be analytic at $Q^2=0$. The method of identifying such
an analytic coupling consists in starting with an appropriate
ansatz for the related beta function and reconstructing from it the
coupling $a(Q^2)$ by solving 
the renormalization group equation (RGE).\footnote{
In another context, an all orders beta function 
for non-supersymmetric Yang-Mills 
theories was proposed in Ref.~\cite{Ryttov:2007cx}, inspired by the 
Novikov-Shifman-Vainshtein-Zakharov beta function of 
${\cal N}=1$ supersymmetric gauge theories \cite{Novikov:1983uc}.}
Such an approach (cf.~Ref.~\cite{Raczka}) is natural because ITEP-OPE 
condition can be implemented in this approach in a very simple way, 
by requiring that beta function $\beta(a)$ as a function of the coupling 
$a$ be analytic in point $a=0$, and have there the pQCD Taylor expansion
$\beta(a) = - \beta_0 a^2 (1 + c_1 a + O(a^2))$ where $\beta_0$
and $c_1=\beta_1/\beta_0$ are universal. 
However, having the ITEP-OPE condition easily
implemented in this way, it turns out to be very difficult to 
find a beta function which gives simultaneously an analytic 
coupling $a(Q^2)$ (i.e., analytic in the complex $Q^2$-plane with the
exception of the negative semiaxis) and which gives high
enough value $r_{\tau} \approx 0.20$ (i.e., compatible with
the experimental measurements). In order to obtain analyticity 
of $a(Q^2)$, we are led to restrict ourselves to
certain classes of beta functions. However, the obtained
values of $r_{\tau}$ turn out to be significantly too low
unless the beta functions are further modified in a peculiar,
perhaps intriguing, manner.

In this letter, in Sec.~\ref{sec:ans1} we motivate the
first class of beta functions which lead to the analyticity
of $a(Q^2)$ while respecting the ITEP-OPE condition.
In Sec.~\ref{sec:ans2} we modify these beta functions in such a way 
as to obtain the correct value of $r_{\tau}$ while 
maintaining the analyticity and the ITEP-OPE condition.
Section \ref{sec:summ} summarizes our results and outlines the prospects 
of further phenomenological applications of the obtained models. 

\section{Beta function ans\"atze for ITEP-OPE and analyticity}
\label{sec:ans1}

The renormalization group equation (RGE)
\be
Q^2 \frac{d a(Q^2)}{d Q^2} = \beta( a(Q^2) ) \ ,
\label{RGE}
\ee
determines the running coupling
$a(Q^2)$ at (in general complex) $Q^2$ once an initial condition
$a(Q^2_{\rm in}) = a_{\rm in}$ is imposed. 
We will impose the initial condition in the present anQCD versions 
at the scale $Q^2_{\rm in}$ of the $3 \to 4$ active quark flavor threshold;
we choose $Q^2_{\rm in} = (3 m_c)^2$ ($\approx 14.5 \ {\rm GeV}^2$). 
The value of $a_{\rm in}=a(Q^2_{\rm in})$ is obtained 
as usual in perturbative QCD (pQCD).
The analytic coupling we get is valid for $n_f=3$ (the three 
active quarks $u$, $d$ and $s$ being almost massless),  
and for higher energies the standard pQCD couplings can be used because 
our versions of anQCD, at such high energies, practically merge with the 
pQCD due to the ITEP-OPE condition:
\be
\vert a(Q^2) - a_{\rm pt}(Q^2) \vert < (\Lambda^2/Q^2)^n
\label{ITEP-OPE}
\ee
at $Q^2 \gg \Lambda^2 (\sim 0.1 \ {\rm GeV}^2)$ for all positive $n$.  

More specifically, in the renormalization
scheme (RSch) dictated by the expansion of our beta function
$\beta(a)$ in powers of $a$ (i.e., the parameters $c_n \equiv
\beta_n/\beta_0$, for $n \geq 2$), we will require that our $a(Q^2)$ at 
$Q=3 m_c$ achieves such a value $a((3 m_c)^2)=a_{\rm in}$ which leads to 
the value $a(M_Z^2;{\overline {\rm MS}}) =0.119/\pi$ once we (exactly)
change the RSch at $Q= 3 m_c$ to ${\overline {\rm MS}}$ and run
the coupling to $Q=M_Z$ with 
${\overline {\rm MS}}$ perturbative RGE.
The latter running is performed at four-loop level,
taking the RGE thresholds $n_f=3 \mapsto n_f=4$ at
$Q=3 m_c$ and $n_f=4 \mapsto n_f=5$ at $Q=3 m_b$, using
the procedure of Ref.~\cite{CKS} with three-loop threshold matching
conditions (for two-loop matching conditions, 
cf.~Refs.~\cite{BW,RS,LRV}). We note that
the value $a(M_Z^2;{\overline {\rm MS}}) \approx 0.119/\pi$ is obtained
by application of pQCD evaluations to QCD observables of 
higher energies ($|Q^2| \stackrel{>}{\sim} 10 \ {\rm GeV}^2$).

With such a fixing of the initial condition, 
integration of RGE (\ref{RGE}) in the complex $Q^2$-plane
can be made more transparent by introducing the new complex
variable $z= \ln(Q^2/\mu^2_{\rm in})$, with $\mu_{\rm in}$
being a fixed scale; we chose $\mu_{\rm in} = 3 m_c$.
The entire complex $Q^2$-plane (the first sheet)
then corresponds to the complex $z$ stripe:
$- \pi \leq {\rm Im}(z) < + \pi$.
The complex $Q^2$-plane 
${\mathbb{C}}\backslash (-\infty,0]$
where $a(Q^2)$ has to be analytic corresponds to the
complex $z$ stripe $- \pi < {\rm Im}(z) < + \pi$,
while the Minkowskian semiaxis $Q^2 \leq 0$ corresponds
to ${\rm Im}z = - \pi$; the point $Q^2=0$ corresponds to
$z= - \infty$, and $Q^2= (3 m_c)^2$ to $z=0$.
Using the notation $a(Q^2) \equiv F(z)$, RGE (\ref{RGE})
can be rewritten in the form
\be
\frac{d F(z)}{d z} = \beta(F(z)) \ ,
\label{RGEz}
\ee
in the semi-open stripe $- \pi \leq {\rm Im}(z) < + \pi$,
and requiring for the analyticity of $a(Q^2)$ in
the $Q^2$ sector ${\mathbb{C}}\backslash (-\infty,0]$ equivalently the analyticity
of $F(z)$ in the open $z$-stripe $- \pi < {\rm Im}(z) < + \pi$
($\Rightarrow \partial F/\partial {\overline z} = 0$).
If we write $z=x + i y$, and $F=u + i v$, RGE (\ref{RGEz})
can be rewritten in term of real functions $u$, $v$ and
real variables $x$, $y$
\ba
\frac{\partial u(x,y)}{\partial x} &=& {\rm Re} \beta (u + i v) \ ,
\quad 
\frac{\partial v(x,y)}{\partial x} = {\rm Im} \beta (u + i v) \ ,
\label{RGEx}
\\
\frac{\partial u(x,y)}{\partial y} &=& - {\rm Im} \beta (u + i v) \ ,
\quad 
\frac{\partial v(x,y)}{\partial y} = {\rm Re} \beta (u + i v) \ .
\label{RGEy}
\ea

If $\beta(F)$ in (\ref{RGEz}) is an analytic function of $F$ at $F=0$,
then $a(Q^2)$ fulfills ITEP-OPE condition (\ref{ITEP-OPE}). 
This will be demonstrated in the following lines
by assuming that the ITEP-OPE condition is not fulfiled and 
showing that consequently $\beta(F)$ must be nonanalytic at $F=0$. 
If ITEP-OPE condition (\ref{ITEP-OPE}) is not fulfilled, then
there exists a positive $n_0$ such that
\be
\delta a(Q^2) \equiv a(Q^2) - a_{\rm pt}(Q^2) \approx
\kappa (\Lambda^2/Q^2)^{n_0} 
\label{nonITEP}
\ee
when $Q^2 \gg \Lambda^2$. Due to asymptotic freedom
at such large $Q^2$, $a_{\rm pt}(Q^2)$ is
\be
a_{\rm pt}(Q^2) = \frac{1}{\beta_0 \ln (Q^2/\Lambda^2)} +
{\cal O} \left( \ln \ln (Q^2/\Lambda^2) / \ln^2 (Q^2/\Lambda^2) \right) \ ,
\label{apt1l}
\ee
and the power term can be written as
\be
(\Lambda^2/Q^2)^{n_0} \approx \exp \left( -K/a_{\rm pt}(Q^2) \right) \ ,
\label{powt1}
\ee
where\footnote{
If the terms 
${\cal O} \left( \ln \ln (Q^2/\Lambda^2) / \ln^2 (Q^2/\Lambda^2) \right)$
in Eq.~(\ref{apt1l}) are included, expression $\exp(-K/a_{\rm pt})$
gets replaced by 
$\exp(-K/a_{\rm pt}) (\beta_0 a_{\rm pt})^{- n_0 \beta_1/\beta_0^2}
\left( 1 + {\cal O}(a \ln^2 a) \right)$;
this does not change the argument in the text.}
$K = n_0/\beta_0$. When we apply $Q^2 d/d Q^2$ to relation
(\ref{nonITEP}), and use expression (\ref{powt1}), we obtain
\be
\beta(a(Q^2)) - \beta_{\rm pt}(a_{\rm pt}(Q^2)) \approx
- n_0 \kappa \exp \left( -K/a_{\rm pt}(Q^2) \right) \ .
\label{difbet1}
\ee
Now we can replace $a(Q^2)$ in the first beta function
in Eq.~(\ref{difbet1}) by $a_{\rm pt}(Q^2) + \kappa 
\exp ( -K/a_{\rm pt}(Q^2) )$, due to relations (\ref{nonITEP})
and (\ref{powt1}), and Taylor-expand the $\beta$-function around
$a_{\rm pt}(Q^2)$ ($\not= 0$). This then gives
\be
\beta(a_{\rm pt}) + \kappa \exp ( -K/a_{\rm pt}) 
\frac{d \beta(a)}{d a} {\big |}_{a=a_{\rm pt}}
+ {\cal O} \left( \exp ( -2 K/a_{\rm pt}) \right) =
\beta_{\rm pt}(a_{\rm pt}) -  n_0 \kappa \exp ( -K/a_{\rm pt}) \ .
\label{difbet2}
\ee
Since this relation is valid for small values of $|a_{\rm pt}|$,
the derivative $d \beta (a)/d a$ at $a=a_{\rm pt}$ on 
LHS of Eq.~(\ref{difbet2}) is very small (about 
$-2 \beta_0 a_{\rm pt}$) and can be neglected. 
This means that Eq.~(\ref{difbet2}) can be rewritten 
for small values of $a_{\rm pt}=F$ as\footnote{
Due to a typo, in the published version (JPG37,075001) the
second term on the RHS of Eq.~(\ref{difbet3}) has ``+'' sign
instead of ``-'' sign.}
\be
\beta(F) \approx \beta_{\rm pt}(F) - n_0 \kappa \exp ( -K/F) \ .
\label{difbet3}
\ee
While $\beta_{\rm pt}(F)$ is analytic at $F=0$, the
term $\exp(-K/F)$ is nonanalytic at $F=0$. Therefore,
the non-fulfillement of ITEP-OPE condition (\ref{ITEP-OPE})
implies nonanalyticity of $\beta(F)$ at $F=0$, and this concludes
the demonstration.

In addition, since at small $F$ the beta function
has to respect pQCD, the following condition must be imposed
on it:
\be
\beta(F) = - \beta_0 F^2 \left[ 1 + c_1 F + c_2 F^2 + \cdots \right] \ ,\
\label{c1con}
\ee
where the parameters $\beta_0$ and $c_1 = \beta_1/\beta_0$
are universal; at $n_f=3$ we have $\beta_0=9/4$ and
$c_1=16/9$.
 
A high precision implementation of the numerical integration of 
RGE (\ref{RGEx})-(\ref{RGEy}), e.g., with Mathematica \cite{math},
for various ans\"atze of $\beta(F)$ function
and respecting pQCD condition (\ref{c1con})
and the ITEP-OPE condition [analyticity of $\beta(F)$ in $F=0$]
then indicates that it is in general difficult to obtain
a result $F(z)$ analytic in the entire open stripe
$- \pi < {\rm Im}(z) < + \pi$. In our approach we assume that
the analytic coupling $a(Q^2)$ reflects all the major
analyticity aspects of the space-like observables ${\cal D}(Q^2)$
(such as Adler function, Bjorken sum rules, etc.). 
This means that our $a(Q^2)$ is analytic even at the origin
$Q^2=0$ ($\Leftrightarrow \ z=- \infty$). This condition, in general, implies
\be
a(Q^2) = a_0 + a_1 (Q^2/\Lambda^2) + {\cal O}[ (Q^2/\Lambda^2)^2] \ ,
\label{aQ0}
\ee
where $a_0=a(Q^2=0) = F(z=-\infty) < \infty$. By applying to Eq.~(\ref{aQ0}) 
the RGE derivative $Q^2 (d /d Q^2)$, we can see
that the beta function $\beta(a)=\beta(F)$ then has a Taylor expansion
around the point $a_0$ with the first Taylor coefficient equal to
unity
\be
\beta(F) = 1 \times (F - a_0) + {\cal O}[ (F - a_0)^2 ] \ ,
\label{anQ0a}
\ee
which can be equivalently expressed as
\be
\beta^{\prime}(F) \vert_{F=a_0} = +1 \ .
\label{anQ0b}
\ee
If assuming the analyticity of $a(Q^2)$ at $Q^2=0$
in a more exceptional way
$a(Q^2) = a_0 +  {\cal O}[ (Q^2/\Lambda^2)^n]$ with $n \geq 2$,
this implies the condition $\beta^{\prime}(F) \vert_{F=a_0} = n$;
it turns out that in such cases the RGE-solution $F(z)$ 
has Landau singularities, at ${\rm Im} z = \pm \pi/n$; 
therefore, we discard such a case.

PQCD condition (\ref{c1con}) for the universal parameters
$\beta_0$ and $c_1$, the $Q^2=0$ analyticity condition 
(\ref{anQ0b}), and the ITEP-OPE condition
can then be summarized in the following form of the beta functions:
\be
\beta(F) = - \beta_0 F^2 (1 - Y) f(Y) \vert_{Y \equiv F/a_0} \ ,
\label{ans1}
\ee
where function $f(Y)$ is analytic at $Y=0$ (ITEP-OPE) 
and at $Y=1$ and fulfills the conditions
\ba
f(Y) & = & 1 + (1 + c_1 a_0) Y + {\cal O}(Y^2) \ ,
\label{calfpt}
\\
a_0 \beta_0 f(1) &=& 1 \ .
\label{calfQ0}
\ea
Eq.~(\ref{calfpt}) is the pQCD condition (reproduction of the
universal $c_1$), and Eq.~(\ref{calfQ0})
is the $Q^2=0$ analyticity condition (\ref{anQ0b}).
Under such conditions, and the aforementioned initial condition
at $Q^2 = (3 m_c)^2$, it turns out that certain classes of
functions $f$, upon RGE integration (\ref{RGEz}), do lead to 
analytic coupling $F(z)$. Even more so, the $Q^2=0$ analyticity
condition leads in general to solutions $F(z) = a(Q^2)$ which
have analyticity even on a certain segment of the negative $Q^2$-axis
[$\leftrightarrow {\rm Im}(z) = - \pi$]: $- M_{\rm thr}^2 < Q^2 \leq 0$
[$\leftrightarrow  -\infty < {\rm Re}(z) < x_{\rm thr}$],
$M_{\rm thr}$ being a ``threshold'' mass, i.e., the cut semiaxis in the 
complex $Q^2$-plane is $( -\infty,-M_{\rm thr}^2]$.

For example, when $f(Y)$ is a polynomial or a rational
(i.e., Pad\'e, meromorphic) function, then there exist certain
regions of parameters of these $f(Y)$ functions for which
$F(z)$ is analytic ($\leftrightarrow a(Q^2)$ analytic in the
entire complex $Q^2$-plane with the exception of the cut semiaxis
$(-\infty,-M_{\rm thr}^2]$). This can be also checked and seen by 
analytical integration of RGE (\ref{RGEz}) in such cases
\be
z = G(F) \ , \quad G(F(z)) = \int_{a_{\rm in}}^{F(z)} 
\frac{d {\widetilde F}}{\beta({\widetilde F})} \ .
\label{impl1}
\ee
Namely, when $f(Y)$ is a polynomial or rational function, integral
in Eq.~(\ref{impl1}) can be performed explicitly (analytically).
From such a solution one can see that a pole ($F=\infty$)
is attained on the negative $Q^2$ semiaxis (at $Q^2=-M_{\rm thr}^2 < 0$,
i.e., at $z = x_{\rm thr} - i \pi$), and that other poles and 
singularities would not appear at least for certain range of values
of the free parameters \cite{CKV}. The $Q^2=0$ analyticity condition
(\ref{calfQ0}) turns out to be crucial for such a behavior.  

However, in this approach we encounter a serious problem: virtually
all the choices of the $f(Y)$ functions which fulfill the
aforementioned conditions (\ref{calfpt})-(\ref{calfQ0}) and
whose numerical solution is, at the same time, an analytic function,
lead to too low values of the semihadronic
$\tau$ decay ratio (with $\triangle S=0$): 
$r_{\tau} < 0.16$, while we need $r_{\tau} \approx 0.20$.
The ``leading-$\beta_0$'' (LB) contribution is
$r_{\tau}^{(LB)} < 0.15$ for various classes of beta functions that we 
tried; if it is possible to adjust free parameters in the beta function
ans\"atze in order to increase $r_{\tau}^{(LB)}$ beyond values $0.15$,
Landau singularities of the obtained $F(z)$ [$=a(Q^2)$] appear. 
At first, for all the chosen classes of beta functions, the corrections
beyond LB (bLB) to $r_{\tau}$ were very small ($<0.10$), and the value
$r_{\tau} \approx 0.20$ could not be achieved (some elements of
the $r_{\tau}$ calculation are outlined in the Appendix). 

This problem is partly a reflection of the fact that, when
the analytization of the coupling eliminates the offending
nonphysical cut $0 < Q^2 < \Lambda^2$ of $a_{\rm pt}(Q^2)$,
the quantity $r_{\tau}$ tends to decrease because the aforementioned
cut gave a positive contribution to $r_{\tau}$ \cite{Geshkenbein:2001mn}.

\section{R(tau)-problem: modification of beta function ans\"atze}
\label{sec:ans2}

Since LB contribution $r_{\tau}^{({\rm LB})}$ cannot be
increased further, it appears that the only way to increase
the total calculated $r_{\tau}$ is to increase the 
beyond-the-leading-$\beta_0$ (bLB) contributions: 
NLB, ${\rm N}^2 {\rm LB}$, etc.
A choice of the beta function (\ref{ans1}) in our approach
fixes also the coefficients $c_2$, $c_3$, etc., that appear in the power
expansion (\ref{c1con}) of $\beta(F)$ in powers of $F$. On the other hand,
the coefficient $T_2$ in the third term (${\rm N}^2 {\rm LB}$)
of the expansion of $r_{\tau}$ beyond the LB 
[see Eqs.~(\ref{exprt1}) and (\ref{T2})] contains 
a term $-c_2$; if $c_2$ can be made significantly negative  
($c_2 \ll -1$) by a suitable choice of beta function (\ref{ans1}), 
then $T_2$ and, consequently, ${\rm N}^2 {\rm LB}$ term in expansion 
of $r_{\tau}$ will become significantly positive, increasing thus the 
evaluated value of $r_{\tau}$ (note that coefficient $T_1$ of the 
second, NLB, term is accidentally small, $T_1=1/12$, and independent of
beta function). On the other hand, we do not want to
reduce as significantly the LB contribution when we increase
$T_2$; and the universal $c_1$ coefficient must remain unchanged
during such a modification. 

A modification which achieves the aforementioned effects
is the following:
 \ba
f_{\rm old}(Y) \mapsto f_{\rm new}(Y) &=&
f_{\rm old}(Y) f_{\rm fact}(Y) \ , 
\label{modf}
\\
f_{\rm fact}(Y) &=& \frac{ (1 + B Y^2)}{(1 + (B+K) Y^2)} \ ,
\quad (1 \ll K \ll B) \ .
\label{ffact}
\ea
The modification factor $f_{\rm fact}(Y)$ is chosen in
such a way ($K \ll B$) that, for most of the values of $Y$, it is close 
to one. Therefore, it does not change significantly beta function 
(\ref{ans1}). This means that, if before the modification the LB part of
$r_{\tau}$ was reasonably large (say, $0.14$-$0.15$), it will
not be changed (reduced) very significantly now. PQCD condition
(\ref{calfpt}) will not be modified by such  $f_{\rm fact}(Y)$
because it modifies the expansion coefficients of $\beta(F)$
only at order $\sim F^4$ (i.e., $c_2$) and higher.  
However, since $1 \ll K$, the modification factor
$f_{\rm fact}(Y)$ can decrease the value of $c_2$ significantly
and thus increase significantly the third term in the expansion
of $r_{\tau}$. Numerical investigations indicate that this
is really so, and that, moreover, Landau singularities are not
introduced by such a modification. The latter point can be
understood even by analytical (i.e., explicit) integration of the RGE
in such a case when $f_{\rm old}$ is a polynomial or a rational 
function \cite{CKV}.

The solution, however, comes at a price. The aforementioned
modification increases very significantly the absolute values
of the higher expansion coefficients $c_n$ ($n \geq 4$) of beta 
function. As a consequence, coefficients $|T_n|$ [$\approx c_n/(n-1)$]
in the expansion become very large when $n \geq 4$. This means that
the expansion series for $r_{\tau}$ starts showing signs of divergence
after the first four terms. On the other hand, the behavior of
the first four terms (including ${\rm N}^3 {\rm LB}$) indicates
reasonable behavior (similar is the behavior of asymptotically 
divergent perturbation series in pQCD).

The fact that the values of parameters $|c_2|$, $|c_3|$, etc., are
large does not mean that we are working in a ``wrong'' renormalization
scheme (RSch). The specification of the RSch in terms of coefficients $c_j =
\beta_j/\beta_0$ ($j \geq 2$) is apparently a perturbative concept,
applicable in the regime $|Q^2| \gg \Lambda^2$. It appears that our 
beta function $\beta(F)$ not just fixes a certain set of values
$c_j$ ($j \geq 2$), but it reflects also certain nonperturbative aspects
via its set of zeros and poles in the complex $F$-plane.
For example, the finite value $a_0 = a(Q^2=0)$ is a zero of
the beta function; the function $f_{\rm new}(Y)$ [with $Y=a(Q^2)/a_0
= F(z)/a_0$] has possibly some zeros and/or poles on the real axis
[but not in the interval $Y \in (0,1)$],
and it has two zeros and poles on the imaginary axis close to the
origin [at $F = \pm i B^{-1/2}$ and $F= \pm i (B+K)^{-1/2}$, respectively].
It appears that, while we might be able to go from one set
of values of $c_j$'s to another in this framework, we cannot go
to the ``tame'' pQCD schemes such as ${\overline {\rm MS}}$
or 't Hooft RSch. For example, the 't Hooft RSch ($c_2=c_3= \ldots =0$),
under the assumption of the ITEP-OPE condition, gives us
$\beta(F) = - \beta_0 F^2 (1 + c_1 F)$ and the solution in such a
case violates analyticity, it has namely a Landau cut \cite{Gardi:1998qr}).
Thus it cannot be physically equivalent at $Q^2 \stackrel{<}{\sim} \Lambda^2$
to RSch's of our beta functions. The same is true for
${\overline {\rm MS}}$ RSch, at least in its hitherto known
truncated form.
These considerations lead us to intriguing questions which
may be clarified in the future.
\begin{table}
\caption{Input parameter values of the three considered $\beta$-ans\"atze,
and some resulting values of other parameters: $c_2$, $c_3$ of expansion
(\ref{c1con}), and $a(Q^2)$ at $Q=3 m_c$ and $Q=0$ ($n_f=3$ used).
\label{tabparam} } 
\begin{ruledtabular}
\begin{tabular}{llllllll}
$f_{\rm old}$ & input $f_{\rm old}$ & input $f_{\rm fact}$ & $c_2$ & $c_3$ & $x_{\rm thr}$ & $a\left( (3 m_c)^2 \right)$ & $a(0)$
\\
\hline
P30 & $w_1=1 + i 0.45$ & $K=43.2, B=5000$ & -243.6 & -250.1 & -12.00 & 0.0545 & 0.4596 
\\
P11 & $Y_{\rm pole}=-10.$ & $K=7.0, B=4000$ & -213.3 & -293.9 & -6.44 & 0.0577 & 0.1995
\\
EE & $y_1=0.1, k_1=10., k_2=11.$ & $K=5.27, B=1000$ & $-104.5$ & $-322.7$ & -5.88 & 0.0613 & 0.2370 
\end{tabular}
\end{ruledtabular}
\end{table}
\begin{table}
\caption{The first four terms in expansion (\ref{exprt1})
of $r_{\tau}$ and their sum, in the three considered models. 
In parentheses are the corresponding results for expansion (\ref{exprt2}).
RScl parameter is ${\cal C}=0$; $n_f=3$.
The last column are variations ($\delta$) of the sums
when RScl-parameter ${\cal C}$ increases from $0$ to $\ln(2)$.
\label{tabrtau} } 
\begin{ruledtabular}
\begin{tabular}{lllllll}
$f_{\rm old}$ & $r_{\tau}:$ LB (LO) & NLB (NLO) & 
${\rm N}^2{\rm LB}$ (${\rm N}^2{\rm LO}$) &
${\rm N}^3{\rm LB}$ (${\rm N}^3{\rm LO}$) &
sum (sum) & $\delta$
\\
\hline
P30 & 0.1002 (0.0818) & 0.0005 (0.0100) & 0.0952 (0.1016) & 0.0060 (0.0066) & 0.2018 (0.2000) & $2.5\%(2.7\%)$
\\
P11 & 0.1065 (0.0881) & 0.0006 (0.0111) & 0.0892 (0.0961) & 0.0057 (0.0062) & 0.2020 (0.2015) & $1.5\%(1.7\%)$
\\
EE  & 0.1251 (0.0990) & 0.0007 (0.0147) & 0.0666 (0.0774) & 0.0096 (0.0107) & 0.2020 (0.2017) & $2.4\%(2.7\%)$
\end{tabular}
\end{ruledtabular}
\end{table}
\begin{table}
\caption{Bjorken polarized sum rule (BjPSR) results $d_{\rm Bj}(Q^2)$
in the three considered models, for the sum of the first four terms
in expansion (\ref{Bjres}). RScl parameter is ${\cal C}=0$; $n_f=3$.
In parentheses, the corresponding results for the first four terms of
expansion (\ref{Bjnonres}) are given. In brackets, the corresponding
variations of the results under the RScl variation are given (see the
text for details). The experimentally measured values 
are (Ref.\cite{Deur:2004ti}):
$0.17 \pm 0.07$ for $Q^2=1 \ {\rm GeV}^2$; $0.16 \pm 0.11$ for
$Q^2=2 \ {\rm GeV}^2$; $0.12 \pm 0.05$ for $Q=2.57$ GeV.
\label{tabBj} } 
\begin{ruledtabular}
\begin{tabular}{llll}
$f_{\rm old}$ & $d_{\rm Bj}(Q^2):\  Q=1$ GeV & $Q=\sqrt{2}$ GeV &
$Q=2.57$ GeV 
\\
\hline 
P30 & 0.248 (0.247) [$4.8\%(5.7\%)$] & 0.201 (0.200) [$4.5\%(5.3\%)$] & 0.145 (0.143) [$3.6\%(4.3\%)$] 
\\
P11 & 0.218 (0.224) [$2.1\%(2.5\%)$] & 0.191 (0.194) [$2.1\%(2.4\%)$] & 0.146 (0.146)  [$2.8\%(3.3\%)$]
\\
EE  & 0.215 (0.227) [$3.1\%(4.2\%)$] & 0.188 (0.194) [$2.3\%(2.9\%)$] & 
0.141 (0.141)  [$2.8\%(3.8\%)$] 
\end{tabular}
\end{ruledtabular}
\end{table}

If we choose for $f \equiv f_{\rm new} = f_{\rm old} f_{\rm fact}$ 
in $\beta$-function
(\ref{ans1}) for the part $f_{\rm old}$ of Eq.~(\ref{modf})
simply a polynomial, the LB-part of $r_{\tau}$ remains
low unless the polynomial degree is at least three
(model ``P30'')
\be
{\rm P30:} \qquad f_{\rm old}(Y) = (1 - w_1 Y) (1 - w_2 Y) (1 - w_3 Y) \ .
\label{P30}
\ee
For $f_{\rm old}$ being a cubic polynomial, the number
of free real parameters is four (two in the polynomial, and
$B$ and $K$ in $f_{\rm fact}$). This is so because initially
we have six real parameters [$w_1$, $w_2$, $w_3$, $B$, $K$, $a_0 =a(0)$];
two of them, e.g. $w_3$ and $a_0$, are eliminated by the
$c_1$ condition (\ref{calfpt}) and the $Q^2=0$ analyticity condition
(\ref{calfQ0}).
The two free parameters in the polynomial 
(e.g., two of the three roots) 
can be adjusted in such a way as to get 
the highest possible values of $r_{\tau}^{{\rm (LB)}}$
($\approx 0.13$) with $f_{\rm old}$ alone (i.e., when 
$f_{\rm fact} \mapsto 1$) while still keeping the holomorphy of $F(z)$.
Then the parameters $B$ and $K$ of $f_{\rm fact}$ can be adjusted
so that $r_{\tau} \approx 0.202$, 
the experimentally measured value.\footnote{
The value of $r_{\tau}$ with $\Delta S=0$ and without mass
contributions is $r_{\tau}=0.202 \pm 0.004$; for details
we refer to Ref.~\cite{CKV}; it is extracted from the
ALEPH-measured  \cite{ALEPH2,ALEPH3,ALEPH4} (V+A)-decay ratio
$R_{\tau}(\Delta S=0)$ as in App.~E of Ref.~\cite{CV2},
by eliminating non-QCD contributions and the (small) quark mass effects.
The result here differs slightly from the one of App.~E of Ref.~\cite{CV2}
($0.204 \pm 0.005$) because of the slightly updated value
of $R_{\tau}(\Delta S=0) = 3.479 \pm 0.011$ (Ref.~\cite{ALEPH4})
and an updated value of $|V_{ud}|=0.97418 \pm 0.00027$
(Ref.~\cite{PDG2008}).}
These adjustments still leave us
certain small freedom in fixing the four parameters
[respecting also the condition (\ref{ffact}): $1 \ll K \ll B$]. 
However, the behavior of $F(z)$ changes only little when we vary the
four parameters under such conditions. In Table \ref{tabparam},
first line (model P30), we present some of the results of this model 
for a representative choice of input parameters 
in this case: 
$w_1=1 + i 0.45$ (and $w_2=1 - i 0.45$; as a consequence,
$w_3=-3.817$; $w_j$'s being the tree inverse roots of $f_{\rm old}$);
$K=43.2$, $B=5000$. In Table \ref{tabrtau}, first line, 
we present results for the first four terms of $r_{\tau}$ 
expansion in the approach described in Appendix 
[Eqs.~(\ref{LBrt2}) and (\ref{exprt1})] and their sum; 
in parentheses, the values of the corresponding first four terms 
are given in the case that no
large-$\beta_0$ (LB) resummation is performed.
We can see that the series of $r_{\tau}$ shows marginal convergence 
behavior when the LB-terms are resummed and three additional correction
terms are included [see Eq.~(\ref{exprt1})]. If LB terms are
not resummed, the convergence behavior is worse. Furthemore,
the estimated value of the fifth term is $\approx -2.0$, i.e.,
the series becomes divergent starting with the fifth term.

In Table \ref{tabBj}, first line, we present the results
of the calculation of the BjPSR $d_{\rm Bj}(Q^2)$ in this P30 model 
for various values of the momentum transfer parameter $Q^2$, taking into
account the first four terms and performing LB resummation
[see Eq.~(\ref{Bjres})]; in parentheses, the corresponding
summation of the first four terms without LB resummation is
given [see Eq.~(\ref{Bjnonres})]. The predicted results are within the 
large experimental uncertainties for $d_{\rm Bj}(Q^2)$, except in the 
case $Q^2 = 1 \ {\rm GeV}^2$ where the model predicts by about one 
$\sigma$ higher value. 

If we choose $f_{\rm old}$ to be a meromorphic rational (i.e., Pad\'e)
function, it turns out that already the simplest diagonal Pad\'e
P[1/1] (i.e., ratio of two linear functions of $Y$) can do the job
(model ``P11'')
\be
{\rm P11:} \qquad f_{\rm old} = \frac{ \left(1 - Y/Y_0 \right)}
{ \left( 1 - Y/Y_{\rm pole} \right) } \ .
\label{P11}
\ee
In this case, we have at first five real parameters
($Y_0$, $Y_{\rm pole}$, $B$, $K$ and $a_0$), but two of them,
e.g. $Y_0$ and $a_0$ are eliminated via the
$c_1$-condition and the $Q^2=0$ analyticity condition,
Eqs.~(\ref{calfpt}) and (\ref{calfQ0}).
We can proceed in the same way as in the previous case 
in order to (more or less) 
fix the three free real parameters $Y_{\rm pole}$, $B$, $K$.
The results of a representative choice of these input parameters
are presented in the second line (model P11) of 
Tables \ref{tabparam}-\ref{tabBj}. The zero of 
$f_{\rm old}(Y)$ turns out to be at $Y_0 =0.6874$.
We see that 
the series for $r_{\tau}$ shows reasonably good convergence
behavior in the first four terms. Inclusion of the
fifth term ($\approx -3.7$) destroys the convergence, 
as in P30 case. Furthermore, BjPSR predictions now all lie 
within the one $\sigma$ uncertainties of experimental values.

It turns out that we can choose the function
$f_{\rm old}$ in certain more complicated ways and fulfill
all the imposed conditions. For example, we can
choose it to be a product of exponential function
of type $(\exp(-Y) - 1)/Y$ and its inverse, both of
them rescaled and translated by specific parameters (model ``EE'')
\be
{\rm EE:} \qquad
f_{\rm old}(Y) =  \frac{ \left( \exp[- k_1 (Y - Y_1)] -1 \right) }
{ [ k_1 (Y - Y_1) ] }
\frac{ [ k_2 (Y - Y_2) ] }{ \left( \exp[- k_2 (Y - Y_2)] -1 \right) }
\times {\cal K}(k_1,Y_1,k_2,Y_2) \ ,
\label{EE}
\ee
where the constant ${\cal K}$ gives just the required normalization
$f_{\rm old}(Y=0)=1$.
At first we have seven real parameters ($Y_1$, $k_1$, $Y_2$, $k_2$, $B$,
$K$, and $a_0$); two of them, e.g., $Y_2$ and $a_0$, are eliminated
by conditions (\ref{calfpt}) and (\ref{calfQ0}). We need
$0 < k_1 < k_2$ to get physically acceptable behavior.
It turns out that with $f$ function being that of Eq.~(\ref{EE})
(when $f_{\rm fact} \equiv 1$, i.e., $K=B=0$), 
the value of $r_{\tau}^{\rm (LB)}$ can be increased maximally
to about $0.15$ while keeping $F(z)$ analytic, if
parameter $Y_1$ achieves the value $Y_1 \approx 0.1$. 
Increasing $Y_1$ further
tends to increase the value of $r_{\tau}^{\rm (LB)}$, but
the analyticity of $F(z)$ is destroyed through appearance
of (Landau) singularities within the stripe
$- \pi < {\rm Im}z < + \pi$. The values of $k_1$ and $k_2$
have to be comparatively large and close to each other
if $r_{\tau}^{\rm (LB)}$ is to be kept large.
Parameters $B$ and $K$ of $f_{\rm fact}$ can then
be adjusted so that $r_{\tau} \approx 0.202$ is reproduced.

In the third line (model EE) of Tables \ref{tabparam}-\ref{tabBj} 
we present the results in this case for a
representative choice of input parameters $Y_1$, $k_1$, $k_2$
and $B$ and $K$. 
We see that now the convergence behavior of
the series of the first four terms of $r_{\tau}$ is quite good,
even when the LB terms are not resummed. Inclusion of the
fifth term ($\approx -1.$) destroys the convergence, as in the
previous two models. Furthermore, the results 
of BjPSR agree well with the measured results.

In both Tables \ref{tabrtau} and \ref{tabBj}
we use the renormalization scale (RScl) parameter 
${\cal C}=0$ [cf.~Eqs.(\ref{exprt1})-(\ref{IanC}),
(\ref{exprt2}), (\ref{Bjres})-(\ref{Bjnonres})]. 
If we vary ${\cal C}$ towards smaller values
[${\cal C}= \ln(1/2)$], the results change insignificantly,
except in the case of BjPSR at $Q^2=1 \ {\rm GeV}^2$ in P11 and EE.
If we increase ${\cal C}$ to $\ln(2)$, the results decrease, 
and the percentages of such decrease of $r_{\tau}$
are given in the last column (``$\delta$'') of Table \ref{tabrtau},
and for BjPSR $d_{\rm Bj}(Q^2)$ are given in Table \ref{tabBj}
in brackets.
Only in the case of BjPSR at $Q^2=1 \ {\rm GeV}^2$ in P11 and EE
these percentages mean the variation (decrease) of the result when
${\cal C}$ goes down to $\ln(1/2)$.
In parentheses, the corresponding values are given when the
LB terms are not resummed, cf.~Eqs.(\ref{exprt2}) and (\ref{Bjnonres}). 
If only three terms are included in our calculations,
the variations of the results for $r_{\tau}$ and BjPSR
under the aforementioned variations of RScl significantly increase,
in general to about $10 \%$. 

If we use as the basis of our calculations of $r_{\tau}$
and $d_{\rm Bj}$ the truncated expansions
in powers $a^n$ like Eq.~(\ref{Adlpt1}), 
instead of the truncated expansions
(\ref{Adlpt2}) in logarithmic derivatives ${\widetilde a}_n$
(\ref{tan}), the results turn out to be significantly
more unstable under the variation of RScl. E.g., the
value $\delta$ in Table \ref{tabrtau} in the case P11
changes from $1.5 \% (1.7 \%)$ to $8.2 \% (9.9 \%)$,
and the value of $r_{\tau}$ changes from $0.2020 \pm 0.0031$
($0.2015 \pm 0.0034$) to $0.2815 \pm 0.0232$
($0.2747 \pm 0.0272$). 

\begin{figure}[htb]
\begin{minipage}[b]{.49\linewidth}
 \centering\includegraphics[width=85mm]{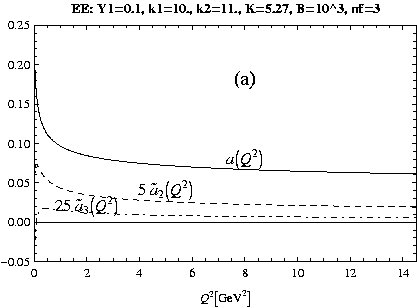}
\end{minipage}
\begin{minipage}[b]{.49\linewidth}
 \centering\includegraphics[width=85mm]{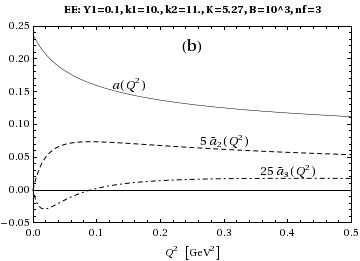}
\end{minipage}
\vspace{-0.4cm}
 \caption{\footnotesize (a) Analytic coupling $a(Q^2)$ (full line)
at positive $0 \leq Q^2 < (3 m_c)^2$, in model EE; included are also higher order
analytic couplings ${\widetilde a}_2(Q^2)$ (dashed line) and 
${\widetilde a}_3(Q^2)$ (dot-dashed line) [cf.~Eqs.~(\ref{tan})],
for better visibility scaled by factors $5$ and $5^2$, respectively;
(b) same as in (a), but at very low $Q^2 > 0$.}
\label{plaEE}
 \end{figure}
%\begin{figure}[htb]
% \includegraphics[height=.40\textheight]{pl3dabsbtEEm2sh.jpg}
% \caption{\footnotesize  Absolute value of $\beta(F(z))$ 
%in model EE as a function of
%$x$ and $y$ (where $z=x + i y$). The only pole is at
%$z_{\rm thr} = x_{\rm thr} \pm i \pi$. The physical sheet is
%$- \pi \leq y < \pi$.} 
%\label{absbetEE}
% \end{figure}
\begin{figure}[htb] %\unitlength=1mm
\centering
\includegraphics[width=100mm]{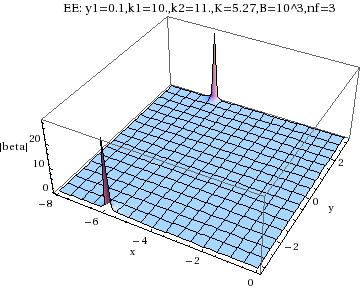}
%width=10.cm,height=6.cm}
\vspace{-0.4cm}
\caption{\footnotesize  Absolute value of $\beta(F(z))$ in model EE as a 
function of $x$ and $y$ (where $z=x + i y$). The only pole is at 
$z_{\rm thr} = x_{\rm thr} \pm i \pi$. The physical sheet is $- \pi \leq y < \pi$.} 
\label{absbetEE}
\end{figure}
\begin{figure}[htb] %\unitlength=1mm
\begin{minipage}[b]{.49\linewidth}
 \centering\includegraphics[width=85mm]{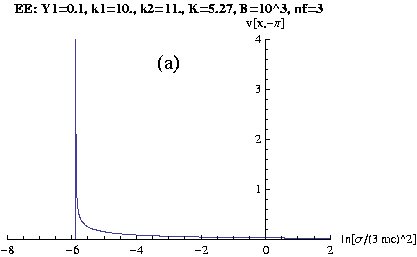}
\end{minipage}
\begin{minipage}[b]{.49\linewidth}
 \centering\includegraphics[width=85mm]{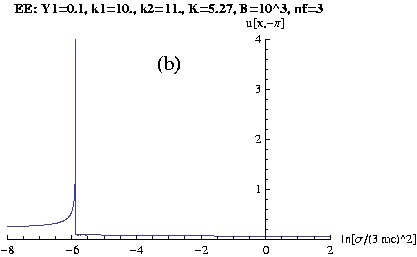}
\end{minipage}
\vspace{-0.4cm}
 \caption{\footnotesize  (a) Imaginary part ${\rm Im} F(z=x - i \pi)
= v(x,-\pi)$ of the analytic coupling $F(z) = a(Q^2)$ in model EE,
as a function of $x$. Here, $v(x,-\pi) = 
{\rm Im} a(Q^2 = - \sigma - i \epsilon) = \rho_1(\sigma)$
is the usual discontinuity function of the analytic coupling,
where $\sigma = \mu_{\rm in}^2 \exp(x)$ ($\mu_{\rm in} = 3 m_c$, 
$m_c=1.27$ GeV).
(b) Same for the real part ${\rm Re} F(z=x - i \pi)= u(x,-\pi)$.}
\label{vuedgeEE}
 \end{figure}

All these results show that model EE is very similar to model P11, but 
significantly different from model P30. 
Further, the threshold values $x_{\rm thr}$ in models EE and P11 
are similar (see Table \ref{tabparam}): $x_{\rm thr}\approx - 6.$;
this corresponds to the threshold mass $Q^2_{\rm thr} = - M^2_{\rm thr}$
for the discontinuity function $\rho_1(\sigma)$ 
with values $M_{\rm thr} = (3 m_c) \exp(x_{\rm thr}/2) \approx 0.2$ GeV.
On the other hand, in model P30, $x_{\rm thr}$ is much more negative:
$x_{\rm thr} \approx - 12.$, corresponding to $M_{\rm thr} \approx 0.01$
GeV. For all these reasons, we will consider models $EE$ and
$P11$ as two viable models of analytic QCD which fulfill the
conditions imposed at the outset of this letter.

In Figs.~\ref{plaEE}(a), (b) we present $a(Q^2)$ and the higher order
couplings ${\widetilde a}_j(Q^2)$ ($j=2,3$) [cf.~Eq.~(\ref{tan})], 
in model EE, as functions of $Q^2$ at low positive 
$Q^2 \leq \mu^2_{\rm in}$.
The Figure indicates strong hierarchy 
$a(Q^2) \gg {\widetilde a}_2(Q^2) \gg {\widetilde a}_3(Q^2) \gg \ldots$
at all positive values of $Q^2$. 
In Fig.~\ref{absbetEE} we present the three-dimensional
image of $|\beta(F(x+i y))|$ as a function of
$x$ and $y$; we can see that there are no
singularities of this function inside the $z$ stripe
$- \pi < y (= {\rm Im}(z) ) < + \pi$; the only
singularity is at the threshold value $z_{\rm thr} =
-5.8754 - i \pi$ which corresponds to
$Q^2 \approx - (0.202)^2 \ {\rm GeV}^2$ on
the negative $Q^2$-axis.
In Figs.~\ref{vuedgeEE} (a), (b) we present the behavior
of the imaginary and real part of the coupling $F$
on the edge $z = x - i \pi$ [i.e., on the negative
$Q^2$ axis: $Q^2 = - \mu_{\rm in}^2 \exp(x)$].
We see the threshold-type behavior at  
$z_{\rm thr} = x_{\rm thr} - i \pi = -5.8754 - i \pi$. 
The fact that these latter curves have no (step-like) discontinuities
at $x \not= x_{\rm thr}$ is an additional numerical indication
that the function $F(z)$ has no singularities within the
stripe $- \pi < {\rm Im z} < \pi$, i.e., no Landau singularities.

\section{Summary}
\label{sec:summ}

We investigated whether it is possible to construct
analytic versions of QCD which obey the ITEP-OPE principle
of no UV-contributions to power term corrections to pQCD $(\Lambda^2/Q^2)^n$
and, at the same time, do not contradict the measured
value of the semihadronic $\tau$ decay ratio $r_{\tau}$ (which is by
far the most precisely measured low energy QCD quantity).
We constructed such models by choosing specific forms for the RGE 
beta-function, and found that the answer is positive:
such theories do exist. However, the obtained solutions 
came at a price, because the obtained series for $r_{\tau}$
show divergent behavior starting with the fifth term of the series.
This was so because we had to introduce poles and zeros of the beta 
function on the imaginary axis relatively close to the origin
(in the complex plane of the coupling), in order to increase
the value of $r_{\tau}$. One model contained a cubic polynomial,
another a simple Pad\'e P[1/1] function, and yet another
model a combination of exponential functions of the type
$(\exp(-Y) - 1)/Y$. The last two models show better apparent
convergence behavior of $r_{\tau}$ (in the first four terms)
and agree well with the (less precisely) measured values
of the Bjorken polarized sum rule at low energies. 
The last two models appear
to be numerically very similar to each other. We intend to use these
two models in the future evaluations of various physical
quantities with the OPE approach. This approach can be
applied with the presented analytic QCD models since the latter
respect the ITEP-OPE philosophy. For example, higher-twist
contributions to the Bjorken polarized sume rule may be substantial.
Such contributions were ignored in the numerical analysis here,
but should eventually be included.

\begin{acknowledgments}
\noindent
This work was supported by FONDECYT Grant No. 1095196 (G.C.),
DFG-CONICYT Project (G.C. and R.K.), and Conicyt (Chile) Bicentenario 
Project PBCT PSD73 (C.V.). 
\end{acknowledgments}

\appendix

\section{Expansions and resummations of observables in analytic QCD}
\label{app1}

Here we refer to and summarize the approach decribed in our previous work
\cite{CV2}. The massless stangeless ($\Delta S=0$)
semihadronic $\tau$ decay ratio
$r_{\tau}$ can be expressed in terms of the current-current
correlation function $\Pi(Q^2)$ (massless, V-V or A-A) as
\be
r_{\tau} = \frac{2}{\pi} \int_0^{m^2_{\tau}} \ \frac{d s}{m^2_{\tau}}
\left( 1 - \frac{s}{m^2_{\tau}} \right)^2 
\left(1 + 2 \frac{s}{m^2_{\tau}} \right) {\rm Im} \Pi(Q^2=-s) \ .
\label{rtauPi}
\ee
This integral can be transformed, via the use
of Cauchy theorem in the $Q^2$-plane\footnote{
In perturbative QCD (pQCD) this use of Cauchy to relation (\ref{rtauPi}) 
is formally not allowed, due to the unphysical (Landau)
cut of $\Pi_{\rm pt} (Q^2)$ along the positive axis 
$0 < Q^2 \leq \Lambda^2$; in pQCD, (\ref{rtauPi}) and
(\ref{rtaucont}) are in principle two different quantities, 
(\ref{rtaucont}) being the preferred one.}
and the subsequent integration by parts, to the contour integral 
\cite{Braaten:1988hc,Beneke:2008ad}
\be
r_{\tau} = \frac{1}{2 \pi} \int_{-\pi}^{+ \pi}
d \phi \ (1 + e^{i \phi})^3 (1 - e^{i \phi}) \
d_{\rm Adl} (Q^2=m_{\tau}^2 e^{i \phi}) \ ,
\label{rtaucont}
\ee
where $d_{\rm Adl} (Q^2) = - d \Pi(Q^2)/d \ln Q^2$ is the
(massless) Adler function whose perturbation expansion is
\ba
d_{\rm Adl.}(Q^2) & = & a + \sum_{n=1}^{\infty} d_n a^{n+1} 
\label{Adlpt1}
\\
& = & 
a + \sum_{n=1}^{\infty} {\widetilde d}_n {\widetilde a}_{n+1} \ .
\label{Adlpt2}
\ea
Here, the coupling parameter $a = a(\mu^2; c_2, c_3,\ldots)$
is at a chosen RScl $\mu^2$ and in a chosen RSch $(c_2, c_3,\ldots)$
($c_n \equiv \beta_n/\beta_0$), as are the coeffficients
$d_n$ and ${\widetilde d}_n$: $d_n=d_n({\cal C}; c_2, \ldots, c_{n-1})$,
${\widetilde d}_n={\widetilde d}_n({\cal C}; c_2, \ldots, c_{n-1})$.
Here, ${\cal C}$ is the dimensionless RScl parameter: 
${\cal C}= \ln(\mu^2/Q^2)$.

The higher order couplings ${\widetilde a}_{n+1}$ appearing in (\ref{Adlpt2}) are
\be
{\widetilde a}_{n+1}(\mu^2) \equiv
\frac{ (-1)^n}{\beta_0^n n!} 
\frac{ \partial^n a(\mu^2)}{\partial (\ln \mu^2)^n} \ ,
\qquad (n=1,2,3, \ldots) \ .
\label{tan}
\ee
The two expansions in (\ref{Adlpt1}) and (\ref{Adlpt2}) are in principle
equivalent (not equivalent in practice, when truncation used), because
of the relations
\ba
{\widetilde a}_2 & = & a^2 + c_1 a^3 + c_2 a^4 + {\cal O}(a^5) \,
\label{ta2}
\\
{\widetilde a}_3 & = & a^3 + (5/2) c_1 a^4 + {\cal O}(a^5) \ ,
\quad {\widetilde a}_4 = a^4 + {\cal O}(a^5) \ , \quad {\rm etc.}
\label{ta3ta4}
\ea
and the consequent relations between $d_n$ and ${\widetilde d}_m$'s
\ba
{\widetilde d}_1 & = & d_1 \ , \quad {\widetilde d}_2 = d_2  - c_1 d_1 \ ,
\label{td1td2}
\\
{\widetilde d}_3 & = & d_3 - (5/2) c_1 d_2 + \left[ (5/2) c_1^2 - c_2 \right] d_1 \ ,
\quad {\rm etc.}
\label{td3}
\ea
The leading-$\beta_0$ contribution (LB, in Refs.~\cite{CV1,CV2}
named leading-skeleton LS) to the massless nonstrange 
ratio $r_{\tau}$ was given in Ref.~\cite{CV2} in Appendix C, 
Eqs.~(C8)-(C11), using results of Refs.~\cite{Neubert,Neubert2}.
It is the contour integration (\ref{rtaucont}) of the
LB-part $d_{\rm Adl}^{\rm (LB)}$ of Adler function
expansion (\ref{Adlpt2}).
While the LB part was written in Refs.~\cite{CV1,CV2}
in terms of the Minkowskian coupling $\tlA_1$
\be
r_{\tau}^{\rm (LB)} = 
\int_0^\infty \frac{dt}{t}\: F_{r}^{\cal {M}}(t) \: 
\tlA_1 (t e^{\cal {\overline C}} m_{\tau}^2) \ ,
\label{LBrt1}
\ee
where ${\cal {\overline C}} = -5/3$, the characteristic function
$F_{r}^{\cal {M}}(t)$ is given in Eqs.~(C10)-(C11) there,\footnote{
A typo appears in the last line of Eq.~(C11) 
of Ref.~\cite{CV2}, in a parenthesis there instead of a term $+3$ should 
be written $+3 t^2$; nonetheless, the correct expression was used in
calculations there.} 
and the Minkowskian (time-like) coupling $\tlA_1(\sigma)$ is related
with the discontinuity (cut) function $\rho_1(\sigma)$
of the coupling parameter $a$ [$\rho_1(\sigma) \equiv
{\rm Im} a(Q^2=-\sigma - i \epsilon)$] in the following way:
\be
\frac{d}{d \ln \sigma} {\tlA}_1(\sigma) =
- \frac{1}{\pi} {\rho}_1(\sigma) \ .
\label{tlA1rho1}
\ee
Since the discontinuity function is $\rho_1(\sigma) =
{\rm Im} F(z)$ for $z=\ln(\sigma/\mu^2_{\rm in}) - i \pi$,
it is obtained as a direct byproduct of the integration of
RGE (\ref{RGEz}). Thefore, it is convenient to express
LB contribution (\ref{LBrt1}) in terms of $\rho_1(\sigma)$
instead of ${\tlA}_1(\sigma)$. This can be obtained from
relation (\ref{LBrt1}) by integration by parts and using
relation (\ref{tlA1rho1})
\be
r_{\tau}^{\rm (LB)} = 
\frac{1}{\pi} \int_0^\infty \frac{dt}{t}\: {\widetilde F}_{r}(t) \: 
\rho_1(t e^{\cal {\overline C}} m_{\tau}^2) \ ,
\label{LBrt2}
\ee
where 
\be
{\widetilde F}_{r}(t) = \int_0^t \frac{dt'}{t'}\: F_{r}^{\cal {M}}(t') \ .
\label{tFtau}
\ee 
Since $F_{r}^{\cal {M}}(t')$ consists of powers of $t'$
and polylogarithmic functions of $t'$ and $1/t'$, it turns out
that integration in (\ref{tFtau}) can be performed analytically.
Explicit expression for ${\widetilde F}_{r}(t)$ will be given in
Ref.~\cite{CKV}. Here we only mention that ${\widetilde F}_{r}(t) \to 1$
when $t \to +\infty$, and that integration in (\ref{LBrt2})
starts at a positive $t_{\rm thr}= (M^2_{\rm thr}/m_{\tau}^2) 
\exp(-{\cal {\overline C}})$, due to the threshold behavior
of $\rho_1(\sigma)$ in our presented models.

A systematic expansion of $r_{\tau}$ beyond the LB can then
be written as $r_{\tau}^{\rm (LB)}$ plus contour integrals of
${\widetilde a}_{n+1}$'s ($n \geq 1$)
\be
r_{\tau} = r_{\tau}^{\rm (LB)} + 
\sum_{n=1}^{\infty} \ T_n I({\widetilde a}_{n+1},{\cal C}) \ ,
\label{exprt1}
\ee
where
\be
I({\widetilde a}_{n+1},{\cal C}) =
\frac{1}{2 \pi} \int_{-\pi}^{+ \pi}
d \phi \ (1 + e^{i \phi})^3 (1 - e^{i \phi}) \
{\widetilde a}_{n+1}(e^{\cal C} m_{\tau}^2 e^{i \phi}) \ ,
\label{IanC}
\ee
${\cal C}$ is an (arbitrary) 
renormalization scale (RScl) parameter ($|{\cal C}| \stackrel{<}{\sim} 1$),
and the coefficients $T_j$ are
\ba
T_1 &=& {\overline T}_1 = {\overline c}_{10}^{(1)} = \frac{1}{12} \ , 
\label{T1}
\\
T_2 & = & {\overline T}_2 + 2 \beta_0 {\cal C} \ {\overline c}_{10}^{(1)} -
(c_2 - {\overline c}_2) \ ,
\label{T2}
\\
T_3 & = & {\overline T}_3 + 3 \beta_0  \ {\cal C} {\overline c}_{10}^{(1)} 
(\beta_0 {\overline c}_{11}^{(2)} + {\overline c}_{10}^{(2)}) +
3 \ {\cal C}  {\overline c}_{10}^{(1)} ( \beta_0^2 {\cal C} - \beta_1)
\nonumber\\
&& + (c_2 - {\overline  c}_2) \left( 
(5/2) c_1 - 3 {\overline c}_{10}^{(1)}
- 3 \beta_0 ({\overline c}_{11}^{(1)} +{\cal C}) \right) 
- (1/2) (c_3 - {\overline c}_3)
\ .
\label{T3}
\ea
The overlines indicate the corresponding quantities which appear in
the ${\overline {\rm MS}}$ RSch with RScl parameter ${\cal C}=0$;
coefficients $c_{ij}^{(k)}$ are determined by the $\beta_0$-expansions
of the perturbation coefficients of the massless Adler function
$d(Q^2)$; for details see Ref.~\cite{CV2}, 
particularly Appendix A.\footnote{In Ref.~\cite{CV2},
notation ${\widetilde \A}_n$ was used instead of ${\widetilde a}_n$,
and ${\widetilde t}_{n+1}$ instead of $T_n$. The power analogs
${\mathcal{A}}_n$ constructed in Refs.~\cite{CV1,CV2}
reduce to powers $a^n$ here because $\beta(a)$ here is
analytic in $a=0$ (as a consequence of ITEP-OPE condition). 
In Eq.~(A18) of Ref.~\cite{CV2} there is a typo, in the first line
the last term there should be 
$- \delta b_{21} 3 ({\overline c}_{11}^{(1)} + {\cal C})$
instead of $- \delta b_{21} 3 {\overline c}_{11}^{(1)}$.
The correct formula was used in the calculations there;
e.g., Eqs.~(89)-(92) in Ref.~\cite{CV2}, 
which follow from Eq.~(A18) there, are correct.} 
In particular, for $n_f=3$: 
$c_{10}^{(1)}=1/12$, $c_{11}^{(1)}=0.691772$;
$c_{10}^{(2)}=-278.673$, $c_{11}^{(2)}=59.2824$.
The ${\rm N}^3 {\rm LB}$ coefficients
${\overline T}_3$ and $T_3$ can now be calculated exactly
because the ${\rm N}^3 {\rm LO}$ perturbative coefficient
${\overline d}_3$ of the massless Adler function is now
known exactly \cite{d3}. In our case ($n_f=3$)
it turns out that ${\overline T}_2=-12.2554$ and
${\overline T}_3=1.55291$. 

Eq.~(\ref{T2}) indicates that ${\rm N}^3 {\rm LB}$ coefficient
$T_2$ becomes large positive [and thus the ${\rm N}^3 {\rm LB}$
term in expansion (\ref{exprt1}) becomes significant positive]
if the beta-coefficient $c_2$ becomes negative: $c_2 \ll -1$.
Futhermore, if $|c_4|$ is large and dominant (as it is in our models),
Eqs.~(\ref{T1})-(\ref{T3}) indicate that $T_4 \approx 
-(1/3) c_4$ and thus $|T_4|$ is large.

If no LB-resummation is performed in $r_{\tau}$ ($\Leftrightarrow$
in $d_{\rm Adl.}$), then $r_{\tau}$ is obtained  by performing
contour integration (\ref{rtaucont}) term-by-term for the sum
(\ref{Adlpt2})
\be
r_{\tau} = I(a,{\cal C}) + 
\sum_{n=1}^{\infty} \ {\widetilde d}_n I({\widetilde a}_{n+1},{\cal C}) \ .
\label{exprt2}
\ee
In practice, we have to truncate sums (\ref{exprt1}) and
(\ref{exprt2}), by including $n_{\rm max} =3$ because only
the first three coefficients $d_n$ ($\Leftrightarrow {\widetilde d}_n$)
are known exactly \cite{d1,d2,d3}.

Bjorken polarized sum rule (BjPSR) $d_{\rm Bj}(Q^2)$ is yet another
QCD observable with measured values (although much less precisely 
than $r_{\tau}$)
at low energies. It can be calculated in a similar way.
Its perturbation expansion can be organized in two ways, like
in Eqs.~(\ref{Adlpt1}) and (\ref{Adlpt2}) for the Adler function.
LB-resummation 
\be
d_{\rm Bj}(Q^2)^{\rm (LB)} = \int_0^{\infty} \frac{dt}{t} \ F_{\rm Bj}(t)
a(t e^{\cal {\overline C}} Q^2) \ ,
\label{dBjLB}
\ee
can be performed with the characteristic function obtained in 
Refs.~\cite{CV1,CV2}
\ba
F_{\rm Bj}(\tau) = 
\left\{
\begin{array}{ll}
\frac{8}{9} \tau \left( 1 - \frac{5}{8} \tau \right) 
& \tau \leq 1 \\ 
\frac{4}{9 \tau} \left( 1 - \frac{1}{4 \tau} \right)
& \tau \geq 1
\end{array}
\right\} \ .
\label{FLSBj3}
\ea
Inclusion of terms beyond the LB gives
\be
d_{\rm Bj}(Q^2) = d_{\rm Bj}(Q^2)^{\rm (LB)} + \sum_{n=1}^{\infty}
(T_{\rm Bj})_n {\widetilde a}_{n+1}(e^{\cal C} Q^2) \ ,
\label{Bjres}
\ee
where coefficients $(T_{\rm Bj})_n$ are analogous to
coefficients $T_n$ of Eqs.~(\ref{T1})-(\ref{T3}),
but this time based on the BjPSR perturbation coefficients
$({\widetilde d}_{\rm Bj})_k$ ($k=1,\ldots, n$) instead of
${\widetilde d}_k$ of Adler function. The perturbation coefficients
$(d_{\rm Bj})_1$ and $(d_{\rm Bj})_2$ are known exactly \cite{LV}, and
for $(d_{\rm Bj})_3$ we use an estimate given in Ref.~\cite{KS}
for $n_f=3$: $({\overline d}_{\rm Bj})_3 \approx 130.$

If LB resummation is not performed, the resulting expression is
\be
d_{\rm Bj}(Q^2) = a(e^{\cal C} Q^2) + \sum_{n=1}^{\infty}
({\widetilde d}_{\rm Bj})_n {\widetilde a}_{n+1}(e^{\cal C} Q^2) \ ,
\label{Bjnonres}
\ee
where the perturbation coefficients $({\widetilde d}_{\rm Bj})_n$
are evaluated at the chosen RScl $\mu^2 = \exp({\cal C}) Q^2$
and in the RSch $(c_2,c_3,\ldots)$ dictated by $\beta$-functions
of our analytic QCD models.

\end{document}